\documentclass{aa}  

\usepackage{graphicx}	
\usepackage{amsmath}	
\usepackage{amssymb}	
\usepackage{xcolor}     
\usepackage{soul}       
\usepackage{subcaption}
\usepackage{float}

\newcommand{\Chitah}{{\sc Chitah}}
\newcommand{\Yattalens}{{\sc YattaLens}}
\newcommand{\Glee}{{\sc Glee}}

\newcommand{\thetaEin}{\theta_{\rm Ein}}
\newcommand{\thetaEinScaled}{\theta^s_{\rm Ein}}
\newcommand{\zl}{z_{\rm l}}
\newcommand{\zs}{z_{\rm s}}

\newcommand{\bd}{\begin{displaymath}}
\newcommand{\ed}{\end{displaymath}}
\newcommand{\be}{\begin{equation}}
\newcommand{\ee}{\end{equation}}
\newcommand{\beaa}{\begin{eqnarray*}}
\newcommand{\eeaa}{\end{eqnarray*}}
\newcommand{\bea}{\begin{eqnarray}}
\newcommand{\eea}{\end{eqnarray}}

\newcommand{\sref}[1]{Section~\ref{#1}}

\newcommand{\fref}[1]{Figure~\ref{#1}}
\newcommand{\fsref}[1]{Figures~\ref{#1}}
\newcommand{\tref}[1]{Table~\ref{#1}}
\newcommand{\eref}[1]{Equation~(\ref{#1})}

\begin{document} 

\title{Survey of Gravitationally-lensed Objects in HSC Imaging (SuGOHI). IV. Lensed quasar search in the HSC survey}

\titlerunning{SuGOHI-q}
\authorrunning{Chan, Suyu, Sonnenfeld et al.}

\author{James~H.~H.~Chan \inst{\ref{epfl},\ref{ntu},\ref{asiaa}},
Sherry~H.~Suyu \inst{\ref{mpa},\ref{tum},\ref{asiaa}},
Alessandro~Sonnenfeld \inst{\ref{leiden},\ref{ipmu}},
Anton~T.~Jaelani \inst{\ref{kindai},\ref{tohoku},\ref{fimipa}},
Anupreeta~More \inst{\ref{ipmu},\ref{iucaa}},
Atsunori~Yonehara \inst{\ref{kyotosu}}, 
Yuriko~Kubota \inst{\ref{kyotosu}},
Jean~Coupon \inst{\ref{ecogia}},
Chien-Hsiu~Lee \inst{\ref{noao}},
Masamune~Oguri \inst{\ref{ipmu},\ref{utokyo},\ref{resceu_utokyo}},
Cristian~E.~Rusu \inst{\ref{naoj}},
\and
Kenneth~C.~Wong \inst{\ref{ipmu},\ref{naoj}}
}

\institute{
Institute of Physics, Laboratory of Astrophysique, \'Ecole Polytechnique F\'ed\'erale de Lausanne (EPFL), Observatoire de Sauverny, 1290 Versoix, Switzerland 
\label{epfl}
\\
\email{hung-hsu.chan@epfl.ch}
\and
Department of Physics, National Taiwan University, 10617 Taipei, Taiwan
\label{ntu}
\and
Academia Sinica Institute of Astronomy and Astrophysics (ASIAA), 11F of ASMAB, No.1, Section 4, Roosevelt Road, Taipei 10617, Taiwan
\label{asiaa}
\and
Max-Planck-Institut f\"ur Astrophysik, Karl-Schwarzschild-Str. 1, 85748 Garching, Germany
\label{mpa}
\and
Physik-Department, Technische Universit\"at M\"unchen, James-Franck-Strae 1, 85748 Garching, Germany
\label{tum}
\and
Leiden Observatory, Leiden University, Niels Bohrweg 2, 2333 CA Leiden, the Netherlands
\label{leiden}
\and
Kavli IPMU (WPI), UTIAS, The University of Tokyo, Kashiwa, Chiba 277-8583, Japan 
\label{ipmu}
\and
Faculty of Science and Engineering, Kindai University, Higashi-Osaka 577-8502, Japan
\label{kindai}
\and
Astronomical Institute, Tohoku University, Aramaki, Aoba, Sendai 980-8578, Japan
\label{tohoku}
\and
Bosscha Observatory,  FMIPA, Institut Teknologi Bandung, Jl. Ganesha 10, Bandung 40132, Indonesia
\label{fimipa}
\and
The Inter-University Center for Astronomy and Astrophysics, Post bag 4, Ganeshkhind, Pune, 411007, India
\label{iucaa}
\and
Department of Physics, Faculty of Science, Kyoto Sangyo University, 603-8555 Kyoto, Japan
\label{kyotosu}
\and
Department of Astronomy, University of Geneva, ch. d'\'Ecogia 16, 1290 Versoix, Switzerland
\label{ecogia}
\and
National Optical Astronomy Observatory, 950 N Cherry Ave, Tucson, AZ 85719, USA
\label{noao}
\and
Department of Physics, University of Tokyo, 7-3-1 Hongo, Bunkyo-ku, Tokyo 113-0033, Japan
\label{utokyo}
\and
Research Center for the Early Universe, University of Tokyo, 7-3-1 Hongo, Bunkyo-ku, Tokyo 113-0033, Japan
\label{resceu_utokyo}
\and
National Astronomical Observatory of Japan, 2-21-1 Osawa, Mitaka, Tokyo 181-8588, Japan
\label{naoj}
\goodbreak
}

\date{\today}

\abstract{
Strong gravitationally lensed quasars provide powerful means to study galaxy evolution and cosmology. 
We use \Chitah\ to hunt for new lens systems in the Hyper Suprime-Cam Subaru Strategic Program (HSC SSP) S16A.
We present 46 lens candidates, of which 3 are previously known. 
Including 2 additional lenses found by \Yattalens, we obtain X-shooter spectra of 6 promising candidates for lens confirmation and redshift measurements.  
We report new spectroscopic redshift measurements for both the lens and source galaxies in 4 lens systems. 
We apply the lens modeling software \Glee\ to model our 6 X-shooter lenses uniformly.
Through our analysis of the HSC images, we find that HSCJ022622$-$042522, HSCJ115252$+$004733, and HSCJ141136$-$010216 have point-like lensed images, and that the lens light distribution is well aligned with mass distribution within $6\deg$.
Thanks to the X-shooter spectra, we estimate fluxes on the Baldwin-Phillips-Terlevich (BPT) diagram, and find that HSCJ022622$-$042522 has a probable quasar source, based on the upper limit of the {\sc Nii} flux intensity. We also measure the FWHM of Ly$\alpha$ emission of HSCJ141136$-$010216 to be $\sim$$254$~km/s, showing that it is a probable Lyman-$\alpha$ emitter.
}


\maketitle


\section{Introduction}
\label{sec:intro}

Strong gravitationally lensed quasars, though very rare, provide powerful means to study both galaxy evolution and cosmology. 
For galaxy evolution, we can study galaxy mass structures and substructures through the use of the positions, shapes, and fluxes of lensed images \citep[e.g.,][]{SuyuEtal12,Dalal&Kochanek02,VegettiEtal12,NierenbergEtal17,GilmanEtal19}. 
For cosmology, measuring time delays between multiple images allows us to determine the time-delay distance and infer the Hubble constant, $H_0$ \citep[e.g.,][]{Refsdal64,CourbinEtal11,SuyuEtal10,SuyuEtal13,BonvinEtal17,ChenEtal19,WongEtal19}.
The Hubble constant is a crucial cosmological parameter that sets the age, size, and critical density of the Universe, and measuring it independently through lensed quasars is  important given the current tensions in its measurement \citep[e.g.,][]{PlanckCollaboration18,RiessEtal19,FreedmanEtal19,WongEtal19}.
Further, quasar microlensing events which are expected to arise frequently in lensed quasars enable us to investigate various astrophysical questions, such as structure of quasar central engine \citep[e.g.,][]{YoneharaEtal98,MineshigeYonehara99,PoindexterEtal08}, mass function of stars in galaxies \citep[e.g.,][]{WyitheEtal00}, and extra-galactic planet detection \citep[e.g.,][]{DaiGuerras18}.

There have been several undertakings to look for them with various surveys.
The Cosmic Lens All-sky Survey \citep[CLASS;][]{MyersEtal03} discovered the largest statistical sample of radio-loud gravitational lenses by obtaining high-resolution images of flat-spectrum radio sources and identifying the ones that showed multiple images.
In the optical, the SDSS Quasar Lens Search
\citep[SQLS; e.g.,][]{OguriEtal06,OguriEtal08,OguriEtal12,InadaEtal08,InadaEtal10,InadaEtal12} has obtained the largest lensed quasar sample to date based on both morphological and color selection of spectroscopically confirmed quasars. 
\cite{JacksonEtal12} further combined the quasar samples from the SDSS and the UKIRT Infrared Deep Sky Survey (UKIDSS) to find small-separation or high-flux-ratio lenses.
Data mining on catalog magnitudes also provides an opportunity to find lensed quasars \citep{AgnelloEtal15,Agnello17,OstrovskiEtal17,WilliamsEtal18}.
\cite{ChanEtal15} built \Chitah\ to inspect image configurations using lens modeling, which was first demonstrated by \cite{MarshallEtal09} who detected lenses in the Hubble Space Telescope ({\it HST}) archival images.

Another systematic approach has been proposed by \cite{KochanekEtal09} where all extended variable sources are identified as potential lenses. Chao et al. (submitted) have built an algorithm using the extent of variable sources in the difference images of the ongoing HSC Transient Survey.
In addition, the recent data releases of {\it Gaia}, with its exceptional resolution, provide an efficient way to find lensed quasars.
One could conduct quasar lens search by looking for multiple detection in Gaia or comparing the flux and position offsets from other surveys \citep[e.g.,][]{LemonEtal17,LemonEtal18,LemonEtal19,DelchambreEtal19}.
Although not specific to lensed quasars, {\sc Space Warps} \citep{MarshallEtal16,MoreEtal16} show that lensed quasars could also be found through citizen science.

We presented a new lens sample as part of the Survey of Gravitationally-lensed Objects in HSC Imaging (SuGOHI) that aims to find lenses at both galaxy- and cluster-mass scales.
Most of the candidates in this paper are classified by \Chitah, and we refer to this corresponding sample of lenses as the SuGOHI lensed quasar sample, or SuGOHI-q. 
The first SuGOHI galaxy-scale sample (SuGOHI-g) is presented in \cite{SonnenfeldEtal18}, with subsequently discovered lenses described in \cite{WongEtal18}. 
In an upcoming paper, we will present a new sample of lenses obtained by looking at clusters of galaxies (SuGOHI-c, Jaelani et al. in prep.)

This paper is organized as follows. In \sref{sec:hsc}, we briefly introduce the HSC survey. The preselection methods are described in \sref{sec:preselection}.
We recap \Chitah's machinery and present the candidates in \sref{sec:trophies}. 
The X-shooter follow-up is described in \sref{sec:xshooter}.
We confirm our lens systems in \sref{sec:new_lens}.
We conclude in \sref{sec:conclusion}.
All images are oriented with North up and East left.

\section{HSC Survey}
\label{sec:hsc}
The Hyper-Suprime Cam (HSC) has 104 science CCDs covering a field of view of $1.5 \deg$ in diameter with a $0.168\arcsec$ pixel scale for the 8.2 m Subaru telescope \citep{MiyazakiEtal18,KomiyamaEtal18,KawanomotoEtal18,FurusawaEtal18}. 
The Hyper Suprime-Cam Subaru Strategic Program (HSC-SSP) Survey consists of three layers (Wide, Deep, and Ultradeep), and the Wide layer is planned to observe a sky area of $\sim 1400 \deg^2$ in five broadband filters ($grizy$) \citep[see details in][]{AiharaEtal18}. We use imaging data from S16A data release covering $456 \deg^2$ from all five bands, $178\deg^2$ of which has full color to the target depth.
The median seeing in the $i$-band is about $0.6\arcsec$. 
The data is reduced using the pipeline {\tt hscPipe} \citep{BoschEtal18}. Although data from the S16A release is not public at the time of working on this project, most of the lens candidates presented in this work are visible in the public data release 1 (PDR1).

\section{Preselection method}
\label{sec:preselection}

Before running \Chitah, we pre-select our targets to speed up the classification.
The beginning sample comes from either the catalogs with possible lens galaxies, or the catalogs with possible quasar sources.
For the possible lens galaxies, we select the luminous red galaxies (LRG) from the SDSS BOSS spectrograph.
For the possible quasar sources, we use the SDSS+WISE photometry.

\subsection{LRGs in BOSS spectroscopy}
One of the reasons that we choose LRGs is that LRGs are massive galaxies which have larger strong lensing cross section ($\propto \sigma^4$, where $\sigma$ is the velocity dispersion). 
Also, LRGs are brighter and more visible at higher redshifts.
Therefore, there is a bias toward the most massive galaxies.

The BOSS survey provides two principle galaxy samples: LOWZ and CMASS. 
The main difference between the two samples is mostly the redshift distribution: LOWZ galaxies are mostly at $z < 0.4$ while CMASS galaxies are mostly in the range $0.4 < z < 0.7$. 
The number of BOSS galaxies with photometry in all five bands of the 2016A data release of HSC is $\sim 43,000$, of which $\sim 9,000$ are from LOWZ and $\sim 34,000$ from CMASS. 
We include one more LRG catalog provided by \cite{KazinEtal10}\footnote{\url{https://cosmo.nyu.edu/~eak306/SDSS-LRG.html}}, which has $\sim 2,000$ objects.

\subsection{QSOs with the SDSS+WISE photometry}
To find lensed quasar systems that do not have lens galaxies identified as LRGs, we further perform photometric selection of quasar candidates from SDSS Data Release 14 \citep{AbolfathiEtal18} by using a non-parametric Bayesian classification method \cite[e.g.,][]{RichardsEtal04} which incorporates a Kernel Density Estimate \citep[KDE;][]{Silverman86}.

SDSS photometric data is taken under substantially worse seeing condition compared to HSC survey data, and images of lensed quasar systems in SDSS data are expected to show extended structure due to foreground lens galaxy, and/or, multiple images of the lensed quasar.
Therefore, in our photometric selection of quasar candidate, we do not take into account any morphological information such as the probability that the object is point-like (which is often used in the selection of unlensed quasars).  Through only photometric data, we classify objects in the photometric catalog into 3 categories: ``star (S)'', ``galaxy (G)'', and ``quasar (Q)''.
For objects with photometric data $\vec{x}$, the probability of an object to be in category $i= \text{S, G, or Q}$ is evaluated from
\be
P( i | \vec{x} ) =
  \frac{P( \vec{x} | i ) P( i ) }{\sum_{i} P( \vec{x} | i ) P( i )}, 
\ee
where $P( \vec{x} | i ) $ and $P( i )$ are the probability density function (PDF) for category ``$i$'' and  the probability that the object is in category $i$, respectively.
To obtain the PDF for any given photometric data $\vec{x}$,
we applied KDE with the following form:
\be
  P ( \vec{x} | i ) = \frac{1}{N_i}
  \sum_{k=1}^{N_i} \frac{1}{\sqrt{2\pi}h}
  \exp \left( -\frac{\left| \vec{x} - \vec{x}_k \right|}{2h^2} \right), 
\ee
where $N_i$ is the number of objects in category $i$, $\vec{x}_k$ is the photometric data of the $k$-th object in the category, and $h$ is a scaling factor of the kernel function.
In our current study, 5 independent colors, {\it u-g}, {\it g-r}, {\it r-i}, {\it i-z} in SDSS photometry and W1-W2 in WISE photometry, are used for photometric data $\vec{x}$, and $h$ is set to be $0.1$ to maximize the classification accuracy.

Here, we use SDSS-DR14 spectroscopic catalog with WISE photometry, which includes spectroscopically confirmed objects (138,055 quasars, 939,101 galaxies, and 187,431 stars). 
Our final target is multiple quasars behind lens galaxy, and an image of such objects are expected to show extended source like morphology. 
Therefore, we have selected \verb|cModelMag| magnitude from SDSS catalog as magnitude of objects, and have not put any constraint on the source extent such as \verb|probPSF| in SDSS catalog. 
Half of them is used as a training data set to construct the PDF, and the remaining half is used as a test data set to evaluate proper
threshold for $P( {\mathrm Q} | \vec{x} )$ to select as many quasar candidates as possible with high classification accuracy. 
It is not easy to estimate the true value of $P( i )$ in the real Universe due to several biases, and we set $P( i )$ based on the spectroscopic sample we used.
While this is simple, our result does not dramatically change in cases when we assume real number of galaxies that are $10$ times larger than the galaxies in the spectroscopic sample.
After several estimations by using the data sets, we set $P( {\mathrm Q} | \vec{x} ) \ge 0.9$ as a threshold for quasar candidate selection.
With this threshold value, we are expected to obtain quasar candidates which includes $\sim 75\%$ of all quasar with $\sim 95\%$ purity (fraction of quasars in all object which classified as ``quasar'').
We evaluate $P( i | \vec{x} )$ for all objects in SDSS-DR14 photometric catalog with WISE photometry, and obtain quasar candidates of $P( {\mathrm Q} | \vec{x} ) \ge 0.9$.
Since this selection method can also find $\ge 80 \%$ of already known lensed quasars in SDSS photometric objects with WISE photometry, 
lens galaxy must not degrade the selection performance seriously. 
In HSC S16A region, the number of quasar candidates is $\sim76,000$ in $\sim3,000,000$ objects.

We include one more QSO catalog with $\sim34,000$ QSOs provided by \cite{BresciaEtal15}\footnote{\url{http://dame.dsf.unina.it/dame_qso.html}} using the Multi Layer Perceptron with Quasi Newton Algorithm (MLPQNA) method to the optical data of SDSS DR10.

\section{Hunting trophies of \Chitah: promising candidates after visual inspection}
\label{sec:trophies}

\Chitah\ \citep{ChanEtal15} is a lens hunter in imaging surveys, based on the configuration of lensed images.
We briefly describe the procedure of \Chitah\ in \sref{subsec:chitah}, and the grading system in \sref{subsec:grading}.
A few additional candidates found through other means are described in \sref{subsec:others}.
In this work, we focus on quad (four-image) systems using \Chitah.

\subsection{\Chitah: strong-gravitational-lens hunter}
\label{subsec:chitah}
The procedure of \Chitah\ is as follows: 
\begin{enumerate}
\item choose two image cutouts, one from bluer bands ($g/r$) and one from redder bands ($z/y$) based on which band has a sharper point-spread function (PSF). 
\item match PSFs in the two selected bands. 
\item disentangle lens light and lensed images according to color information.
\item identify lens center and lensed image positions, masking out the region within $0.5\arcsec$ in radius from the lens center in the lensed arc image 
to prevent misidentifying lensed image positions near the lens center due to imperfect lens light separation. 
\item model the lensed-image configuration with a Singular Isothermal Elliptical (SIE) lens mass distribution.
\end{enumerate}
The outputs of the model are the best-fit parameters of the SIE: 
the Einstein radius ($\thetaEin$), the axis ratio ($q$), the position angle (PA), and the lens center.
The 2-dimensional surface mass density of SIE is expressed as: 
\be
\kappa = \frac{\thetaEin}{2\sqrt{x^2+y^2/q^2}},
\ee
where $(x, y)$ are the coordinates relative to the lens center, along the semi-major and semi-minor axes of the elliptical mass distribution. 
We determine the SIE model parameters by minimising the $\chi^2_{\rm src}$ on the source plane, which is defined as
\be
\label{equ:chi2src}
\chi^2_{\rm src}= \sum_k\frac{|{\bf r}_k-{\bf r}_{\mathrm{model}}|^2}{\sigma_{\mathrm{image}}^2/\mu_k},
\ee
where ${\bf r}_k$ is the respective source position mapped from the
position of lensed image $k$, $\mu_k$ is the magnification at the position of lensed image $k$, 
$\sigma_{\mathrm{image}}$ is chosen to be the pixel scale of HSC ($0.168\arcsec$) as an estimate of the uncertainty, 
and ${\bf r}_{\mathrm{model}}$ is the modeled source position evaluated as a weighted mean of ${\bf r}_k$,
\begin{equation}
\label{equ:src_mod}
{\bf r}_{\mathrm{model}} = \frac{\sum\limits_k{\sqrt{\mu_k}{\bf r}_k}}{\sum\limits_k{\sqrt{\mu_k}}}
\end{equation}
\citep{Oguri10}. Here the index $k$ runs from 1 to 4 for quad systems.
We also use the lens center from the light profile as a prior to constrain the center of the SIE lens mass model.
Therefore, we define the $\chi_{\mathrm{c}}^2$ as
\begin{equation}
\label{equ:chi2c}
\chi_{\mathrm{c}}^2= \frac{ |{\bf x}_{\mathrm{model}}-{\bf x}_{\mathrm{c}}|^2}{\sigma_{\mathrm{c}}^2},
\end{equation}
where ${\bf x}_{\mathrm{c}}$ is the lens center from the light profile,
and ${\bf x}_{\mathrm{model}}$ is the lens center of the SIE model, 
We choose $\sigma_{\mathrm{c}}$ to be the same as $\sigma_{\mathrm{image}}$.  
We further take into account the residuals of the fit to the ``lensed arc'' image from \Chitah.
The difference between the lensed image intensity $I(i,j)$ and the predicted image intensity $I^{\rm P}(i,j)$ is defined as,
\begin{equation}
\label{equ:chi2res}
\chi_{\mathrm{res}}^2 = \sum_{i,j} \frac{{[I(i,j)-I^{\rm P}(i,j)]^2}}{var(i,j)},
\end{equation}
where $i=1...N_{\rm x}$ and $j=1...N_{\rm y}$ are the pixel 
indices in the image cutout of dimensions $N_{\rm x}\times N_{\rm y}$, and $var(i,j)$ is the pixel uncertainty in $I(i,j)$.
Note that $I^{\rm P}$ is obtained from the PSF fitting instead of the lens modeling. Therefore the fluxes of lensed image are not affected by flux anomalies.

The criteria of classification of lens candidates are $\chi^2_{\rm src}+\chi_{\mathrm{c}}^2 < 2 \thetaEin$, 
where $\thetaEin$ is measured in arcsec, and $\chi_{\mathrm{res}}^2 < 100$.
The former criterion allows \Chitah\ to detect lens candidates covering a wide range of $\thetaEin$, 
since typically $\chi^2_{\rm src}$ scales with $\thetaEin$ 
and our tests with mock systems in \citet{ChanEtal15} show that $\chi^2_{\rm src} \lesssim 4$ yields a low false positive rate of $<3\%$.
The latter criterion allows us to further eliminate false positives.
The lens candidates are selected within $0.3\arcsec < \thetaEin < 4\arcsec$.

\subsection{Grading the hunting trophies}
\label{subsec:grading}
We begin with the preselection catalogs, and then we extract stamps ($7\arcsec\times7\arcsec$) from HSC imaging, including the science images in $g, r, i, z,$ and $y$-bands, the variance images and the corresponding PSFs. 
We begin from the LRG catalogs with $\sim45,000$ objects
and from the QSO catalogs with $\sim110,000$ objects.
After \Chitah's classification, we obtain $800$ candidates from LRG catalogs and $3,400$ candidates from QSO catalogs.
The classification rate is $1.5\%$ and $3.1\%$, respectively.
In the first place, J.~H.~H.~C. remove those candidates that are clearly non-lenses but classified lenses by \Chitah, due partly to imperfect PSF matching and partly to nearby objects in cutouts.
This false detection can be improved by pre-selection methods.
After that, we grade them from 0 to 3, according to the following rule:
\renewcommand{\labelitemi}{\textbullet}
\begin{itemize}
    \item 3: almost certainly a lens 
    \item 2: probably a lens 
    \item 1: possibly a lens 
    \item 0: non-lens 
\end{itemize}
Typical aspects taken into consideration in grading are the residual from lens removal and the positions of possible lensed images.
Nine coauthors independently graded each candidate, assigning a score between 0 and 3 with an interval 0.5, similar to  \cite{SonnenfeldEtal18}.
We list our $46$ candidates with grades~$\geq 1.5$ in \tref{tab:cands}, and show them in \fref{fig:gallery}. 
Most of our candidates from LRG catalogs are also found by \Yattalens\ and have been presented in \cite{SonnenfeldEtal18}.
We notice that the LRG pre-selection tends to provide mostly extended source.
In the QSO pre-selection, there are some clear point sources.
The scaled Einstein radii ($\thetaEinScaled$) are listed in column 4 of \tref{tab:cands}:
\be
\thetaEinScaled = \thetaEin\sqrt{\frac{2q^2}{1+q^2}}.
\ee

\subsection{Other candidates}
\label{subsec:others}
\begin{figure*}
	\centering
	\begin{subfigure}{1.0\textwidth} 
	\centering
	    \subcaption[short for lof]{Candidates from preselected LRGs}
		\includegraphics[width=\textwidth]{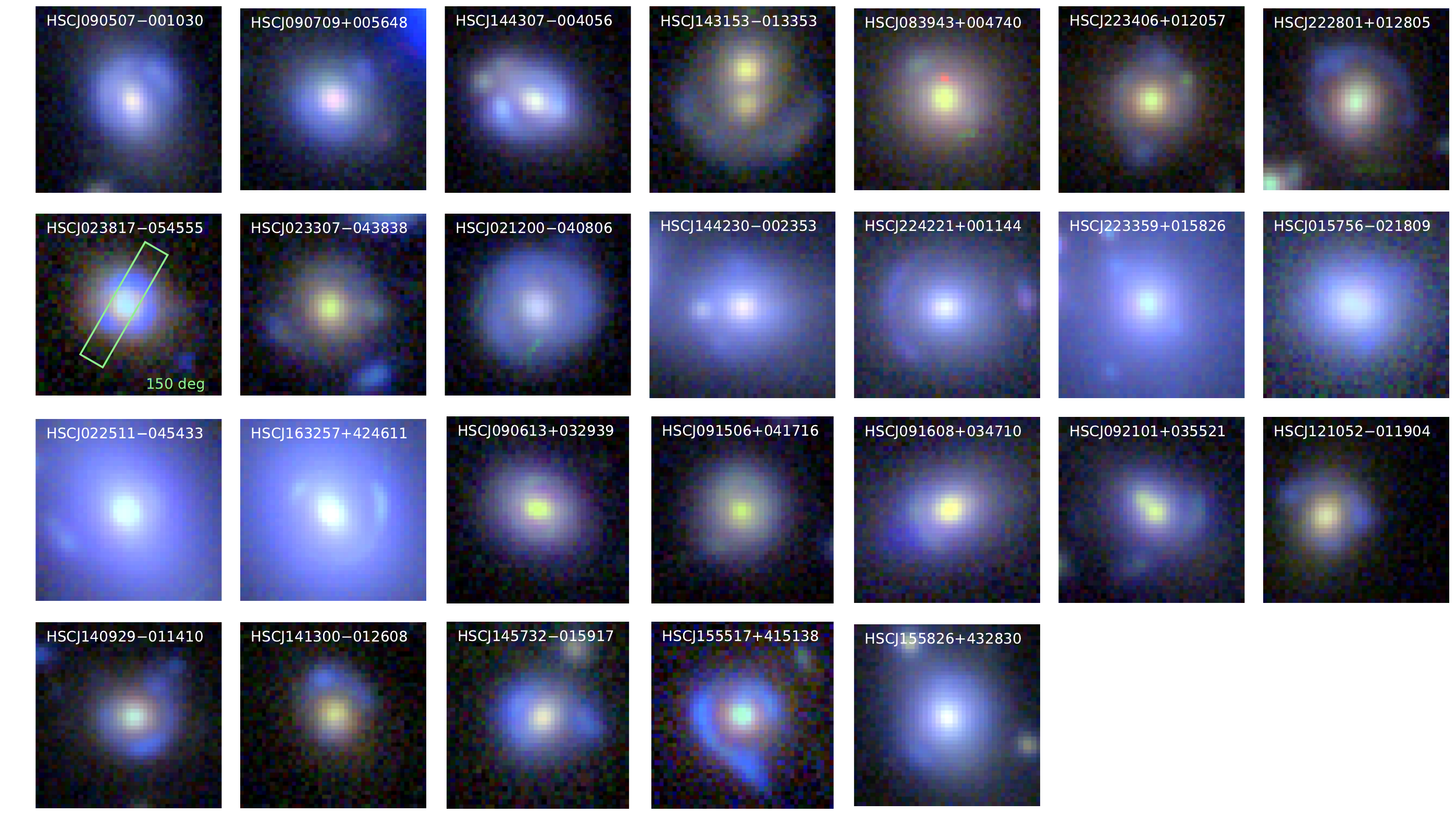}
	\end{subfigure}
	\begin{subfigure}{1.0\textwidth} 
	    \subcaption[short for lof]{Candidates from preselected QSOs}
		\includegraphics[width=\textwidth]{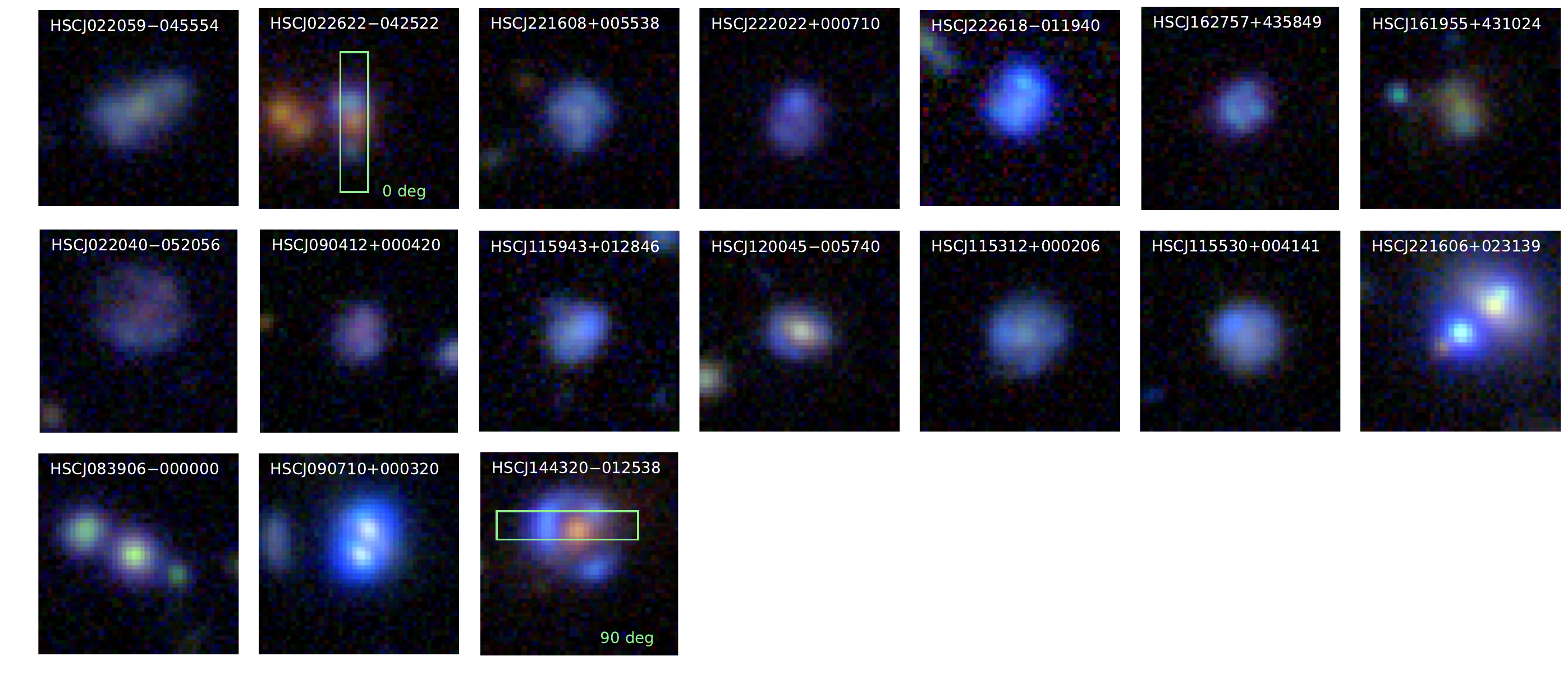}
	\end{subfigure}
	\begin{subfigure}{0.714\textwidth} 
	    \subcaption[short for lof]{Other candidates}
		\includegraphics[width=\textwidth]{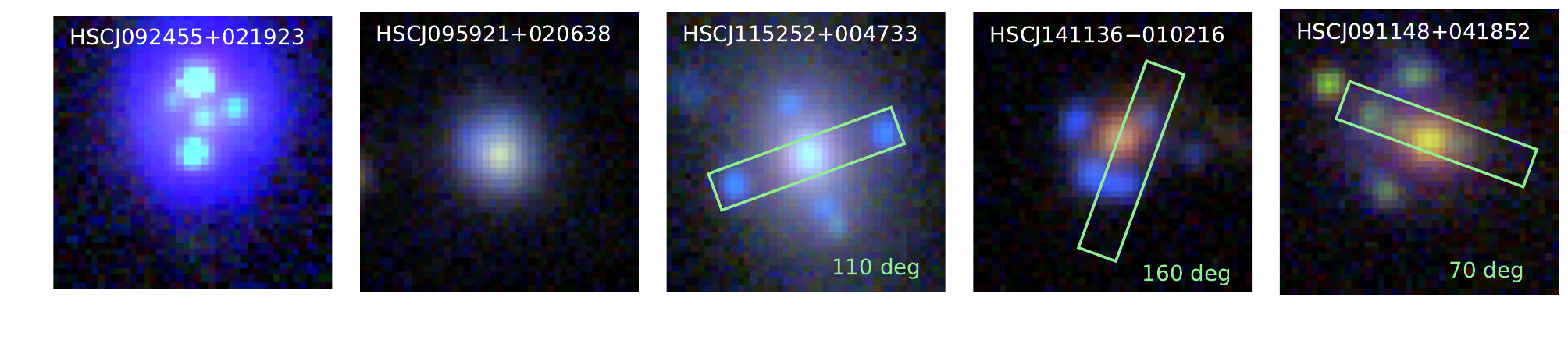}
	\end{subfigure}
	\caption{
	Lens candidates classified by \Chitah\ from preselected LRGs (a) and QSOs (b). 
	We list in (c) other candidates whose origins are noted in \tref{tab:cands}.
	Each $riz$ image cutout is $7\arcsec\times7\arcsec$.
	The green box indicates the position of the slit used during the X-shooter spectroscopic observation.
	} 
	\label{fig:gallery}
\end{figure*}

There are three known lenses in the HSC S16A footprint found again by \Chitah: HSCJ092455$+$021923 \citep{InadaEtal12}, HSCJ095921$+$020638 \citep{AnguitaEtal09}, and HSCJ115252$+$004733 \citep{MoreEtal17}\footnote{We observed HSCJ115252$+$004733 again to determine the nature of its source.}.
We include two new lens candidates, HSCJ091148$+$041852 and HSCJ141136$-$010216, found by \Yattalens\ but missed in \Chitah's classification\footnote{HSCJ091148$+$041852 is preselected from {\sc Space Warps} in the HSC survey (Sonnenfeld et al. in prep.). \Chitah\ missed HSCJ141136$-$010216 since the lensed images are too faint.}, for X-shooter spectroscopic follow-up. 
We present these candidates in \fref{fig:gallery}(c).

\begin{table*}[]
    \begin{center}
\begin{tabular}{lrrccll}
Name &  R.A. [$\deg$] &  Dec [$\deg$] & $\thetaEinScaled$ & Grade & preselection & comment \\
\hline
HSCJ095921$+$020638 &    149.841 &      2.111 &    0.69\arcsec & 3.0 & - & \cite{AnguitaEtal09} \\
HSCJ115252$+$004733$^\dagger$ &    178.218 &      0.793 &    1.39\arcsec & 3.0 & - & \cite{MoreEtal17} \\
HSCJ090507$-$001030 &    136.281 &   $-$0.175 &    1.23\arcsec & 2.1 & LRG: CMASS & \cite{SonnenfeldEtal18} \\
HSCJ090709$+$005648 &    136.790 &      0.947 &    1.32\arcsec & 2.4 & LRG: CMASS & \cite{SonnenfeldEtal18} \\
HSCJ144307$-$004056 &    220.780 &   $-$0.682 &    1.03\arcsec & 2.1 & LRG: CMASS & \cite{SonnenfeldEtal18} \\
HSCJ143153$-$013353 &    217.973 &   $-$1.565 &    2.78\arcsec & 1.8 & LRG: CMASS & \cite{SonnenfeldEtal18} \\
HSCJ083943$+$004740 &    129.929 &      0.795 &    1.40\arcsec & 1.7 & LRG: CMASS & \cite{SonnenfeldEtal18} \\
HSCJ223406$+$012057 &    338.529 &      1.349 &    1.24\arcsec & 1.6 & LRG: CMASS & \cite{SonnenfeldEtal18} \\
HSCJ222801$+$012805 &    337.008 &      1.468 &    1.60\arcsec & 2.6 & LRG: CMASS & \cite{SonnenfeldEtal18} \\
HSCJ023817$-$054555$^\dagger$ &     39.574 &   $-$5.765 &    0.92\arcsec & 2.8 & LRG: CMASS & \cite{SonnenfeldEtal18} \\
HSCJ023307$-$043838 &     38.279 &   $-$4.644 &    1.65\arcsec & 1.8 & LRG: CMASS & - \\
HSCJ021200$-$040806 &     33.004 &   $-$4.135 &    1.59\arcsec & 1.5 & LRG: CMASS & \cite{SonnenfeldEtal18} \\
HSCJ144230$-$002353 &    220.629 &   $-$0.398 &    1.17\arcsec & 1.6 & LRG: LOWZ & - \\
HSCJ224221$+$001144 &    340.590 &      0.196 &    1.41\arcsec & 2.9 & LRG: LOWZ & \cite{SonnenfeldEtal18} \\
HSCJ223359$+$015826 &    338.500 &      1.974 &    0.85\arcsec & 1.5 & LRG: LOWZ & \cite{SonnenfeldEtal18} \\
HSCJ015756$-$021809 &     29.486 &   $-$2.303 &    1.07\arcsec & 1.8 & LRG: LOWZ & \cite{SonnenfeldEtal18} \\
HSCJ022511$-$045433 &     36.296 &   $-$4.909 &    1.53\arcsec & 2.1 & LRG: LOWZ & - \\
HSCJ163257$+$424611 &    248.241 &     42.770 &    1.64\arcsec & 3.0 & LRG: \cite{KazinEtal10} & - \\
HSCJ090613$+$032939 &    136.555 &      3.494 &    0.94\arcsec & 1.6 & LRG: CMASS & \cite{SonnenfeldEtal18} \\
HSCJ091506$+$041716 &    138.776 &      4.288 &    1.23\arcsec & 1.5 & LRG: CMASS & \cite{SonnenfeldEtal18} \\
HSCJ091608$+$034710 &    139.036 &      3.786 &    1.15\arcsec & 2.9 & LRG: CMASS & \cite{SonnenfeldEtal18} \\
HSCJ092101$+$035521 &    140.256 &      3.923 &    1.21\arcsec & 2.6 & LRG: CMASS & \cite{SonnenfeldEtal18} \\
HSCJ121052$-$011904 &    182.718 &   $-$1.318 &    1.16\arcsec & 2.6 & LRG: CMASS & - \\
HSCJ140929$-$011410 &    212.374 &   $-$1.236 &    1.24\arcsec & 3.0 & LRG: CMASS & \cite{SonnenfeldEtal18} \\
HSCJ141300$-$012608 &    213.250 &   $-$1.436 &    1.13\arcsec & 2.8 & LRG: CMASS & \cite{SonnenfeldEtal18} \\
HSCJ145732$-$015917 &    224.386 &   $-$1.988 &    1.20\arcsec & 3.0 & LRG: CMASS & \cite{SonnenfeldEtal18} \\
HSCJ155517$+$415138 &    238.824 &     41.861 &    1.31\arcsec & 3.0 & LRG: CMASS & \cite{SonnenfeldEtal18} \\
HSCJ155826$+$432830 &    239.611 &     43.475 &    1.41\arcsec & 2.0 & LRG: CMASS & \cite{SonnenfeldEtal18} \\
HSCJ092455$+$021923 &    141.233 &      2.323 &    0.84\arcsec & 3.0 & - & \cite{InadaEtal03} \\
HSCJ022059$-$045554 &     35.249 &   $-$4.932 &    1.02\arcsec & 1.6 & QSO: SDSS+WISE & - \\
HSCJ022622$-$042522$^\dagger$ &     36.593 &   $-$4.423 &    0.81\arcsec & 2.6 & QSO: SDSS+WISE & - \\
HSCJ221608$+$005538 &    334.036 &      0.927 &    0.65\arcsec & 2.6 & QSO: SDSS+WISE & - \\
HSCJ222022$+$000710 &    335.095 &      0.120 &    0.62\arcsec & 1.7 & QSO: SDSS+WISE & - \\
HSCJ222618$-$011940 &    336.576 &   $-$1.328 &    0.57\arcsec & 1.6 & QSO: SDSS+WISE & - \\
HSCJ162757$+$435849 &    246.988 &     43.980 &    0.52\arcsec & 1.8 & QSO: SDSS+WISE & - \\
HSCJ161955$+$431024 &    244.982 &     43.173 &    0.77\arcsec & 1.7 & QSO: SDSS+WISE & - \\
HSCJ022040$-$052056 &     35.168 &   $-$5.349 &    0.96\arcsec & 1.7 & QSO: SDSS+WISE & - \\
HSCJ090412$+$000420 &    136.051 &      0.072 &    0.55\arcsec & 1.6 & QSO: SDSS+WISE & - \\
HSCJ115943$+$012846 &    179.931 &      1.479 &    0.57\arcsec & 1.8 & QSO: SDSS+WISE & - \\
HSCJ120045$-$005740 &    180.189 &   $-$0.961 &    1.57\arcsec & 1.9 & QSO: SDSS+WISE & - \\
HSCJ115312$+$000206 &    178.304 &      0.035 &    0.58\arcsec & 1.6 & QSO: SDSS+WISE & - \\
HSCJ115530$+$004141 &    178.879 &      0.695 &    0.64\arcsec & 1.6 & QSO: \cite{BresciaEtal15} & - \\
HSCJ221606$+$023139 &    334.026 &      2.528 &    0.58\arcsec & 1.8 & QSO: \cite{BresciaEtal15} & - \\
HSCJ083906$-$000000 &    129.777 &   $-$0.000 &    1.84\arcsec & 1.8 & QSO: SDSS+WISE & - \\
HSCJ090710$+$000320 &    136.794 &      0.056 &    0.66\arcsec & 1.8 & QSO: SDSS+WISE & - \\
HSCJ144320$-$012538$^\dagger$ &    220.836 &   $-$1.427 &    1.02\arcsec & 2.7 & QSO: SDSS+WISE & - \\
HSCJ091148$+$041852$^\dagger$ &    137.954 &      4.315 & - & - & - & found by \Yattalens \\
HSCJ141136$-$010216$^\dagger$ &    212.902 &   $-$1.038 & - & - & - & found by \Yattalens \\
\end{tabular}
\end{center}

    \caption{
    Lens candidates from \Chitah.
    Column 4 lists $\thetaEinScaled$ modelled by \Chitah. 
    Column 5 shows the average grades of nine coauthors.
    We include the last two candidates found by \Yattalens.
    We highlight those having X-shooter spectra with $\dagger$. 
    }
    \label{tab:cands}
\end{table*}

\section{X-shooter spectroscopic follow-up}
\label{sec:xshooter}
To discern the nature of the lensed candidates, we use the ESO VLT facility with the X-shooter spectrograph \citep{VernetEtal11}. 
The main goal of this programme (ESO programme 099.A-0220, PI: Suyu) is to measure the redshifts of lens galaxies and lensed background sources \citep{SonnenfeldEtal19}.
We observe each target in slit mode with 2 Observation Blocks (OBs), except for HSCJ115252$+$004733 that had only 1 OB due to bad weather.
Each OB corresponds to roughly one hour of telescope time, and consists of $10\times285$s exposures obtained in an ABBA nodding pattern, to optimise background subtraction in the near-infrared (NIR) arm. 
Exposure times in the UVB and VIS arms are slightly shorter due to the longer readout time. 
We use slit widths of 1.0\arcsec, 0.9\arcsec and 0.9\arcsec\ in the UVB, VIS and NIR arms respectively, and applied a $2\times2$ pixel binning to the UVB and VIS CCDs. 
We position the slit so that it covered both the centre of the lens galaxy and the brightest feature of the lensed source.
Observations were executed with a seeing FWHM < 0.9\arcsec on target position.

\begin{figure*}
\centering
\includegraphics[scale=0.43]{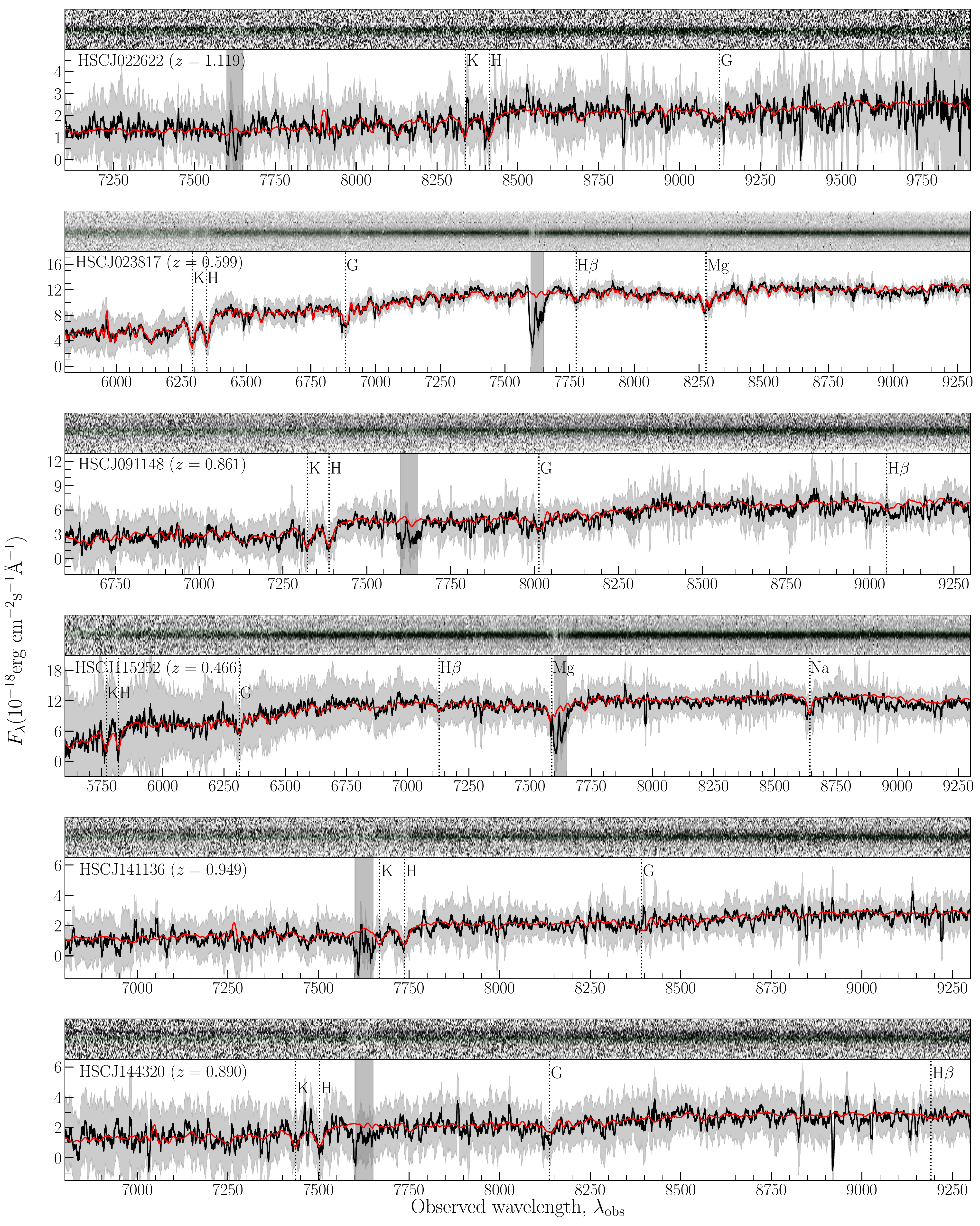}
\caption{
The upper panel in each row shows small cutouts of the 2D spectrum with green shaded region which corresponds to the aperture for 1D spectra extraction of the lens galaxy. The stacked 1D spectrum of the X-shooter lens galaxies are shown with black line with commonly found absorption features indicated by vertical dashed lines. The error on the spectrum is shown with shaded region (grey). The vertical rectangular shaded region (grey) shown in all panels indicate absorption features probably due to telluric contamination. The lens redshift is shown on the upper left in each panel. For comparison, we show a composite luminous red galaxy spectrum from \citet{DobosEtal2012} shifted by the measured redshifts, in red.
The unit of the wavelength is in \AA.
}
\label{fig:specz_lens}
\end{figure*}

\begin{figure*}
\centering
\includegraphics[scale=0.8]{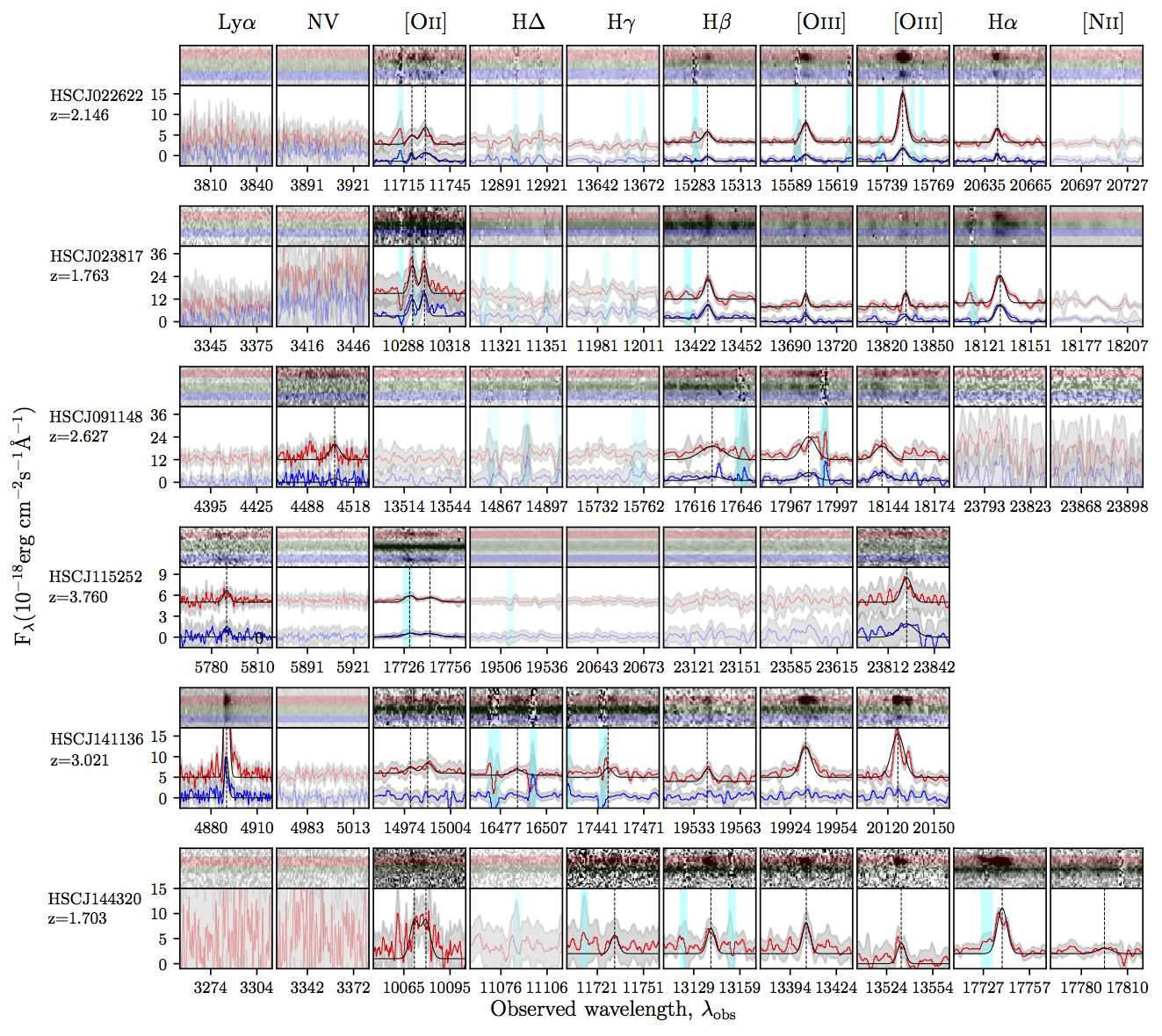}
\caption{
The upper panel in each row shows small cutouts of the 2D spectrum with three shaded regions (red, green, blue) which correspond to the apertures for 1D spectra extraction of lensed source, lens galaxy, and lensed counterpart, respectively. The stacked 1D spectrum of the X-shooter lensed source and its counterpart are shown with red and blue lines, respectively, in the lower panel, with corresponding emission lines marked by the vertical dashed lines and labelled above the panels. The error on the spectrum are shown with shaded region (grey). The black lines show Gaussian fits to the emission lines. The vertical shaded region (cyan) shown in all panels indicates absorption feature due to telluric contamination. The semi-transparent panels show the locations of common emission lines, which are not detected, for the given source redshift. 
The unit of the wavelength is in \AA.
}
\label{fig:specz_src}
\end{figure*}

We reduce the two-dimensional (2D) spectra to one-dimensional (1D) by processing the raw data using \texttt{ESO Reflex} software ver.~2.9.0 combined with the X-shooter pipeline recipes ver.~3.1.0 \citep{FreudlingEtal2013}. In general, the pipeline recipes perform standard bias subtraction, flat-fielding of the raw spectra, and wavelength calibration. Cosmic rays are removed using \texttt{LACosmic} \citep{Dokkum2001}. We calibrate the flux based on spectroscopic standard star.  For further data processing and analysis, we use standard IRAF tools. We stack each 2D single-exposure spectrum of 2 OBs, and produce 1D spectra using an extraction aperture in all three arms. The flux errors are calculated using error propagation from the raw image till extracting the 1D spectra. 
In total, we observed 6 candidates which have probable point-like sources and the 1D spectra of the lens galaxies and lensed sources are shown in \fsref{fig:specz_lens}
and \ref{fig:specz_src}. The extraction apertures for 1D spectra are shown by the red, blue, and green areas in the 2D spectrum for the lensed source, its counterpart, and the lens galaxy, respectively, in \fref{fig:specz_src}.

\section{New lens systems}
\label{sec:new_lens}
After inspecting the spectra of the 6 candidates, we confirm that 5 of them have the same spectra of the lensed source and its counterpart, except for HSCJ144320$-$012538.
However, this object has clear lensed feature as shown in \fref{fig:gallery}, even though we do not obtain the spectrum of its counter image, since it is too faint.

The X-shooter spectra allow us to measure the redshifts of both the lens and source galaxies.
To do so, we smooth the stacked 2D spectrum with a box kernel of 4 \AA \ width, and fit Gaussian profiles on G, H, and K lines for the lens galaxies and detectable emission lines for the lensed sources. 
We list the measurement in \tref{tab:xshooter}.
Evidently, the bluer lensed features are at higher redshifts compare to the main galaxies.
We note that our redshift measurements of HSCJ115252$+$004733 and HSCJ023817$-$054555 are consistent with those in \cite{MoreEtal17} and \cite{SonnenfeldEtal19}, respectively.

\begin{table*}
\begin{center}
\begin{tabular}{cllccrccrcl}
Name & $\zl$ & $\zs$ & $n_\text{S\'ersic}$ & $q_{\rm light}$ & $PA_{\rm light}$ & $\thetaEin$ & $q$ & $PA$ & $\thetaEinScaled$ & comment \\
     &       &       &                     &                 &   [$\deg$]       & [$\arcsec$] &     &[$\deg$]&  [$\arcsec$]    &         \\
\hline
HSCJ022622$-$042522 & $  1.119^*$ & $  2.146^*$ & $  9.13$ & $  0.85$ & $ 10.28$ & $  1.28$ & $  0.56$ & $ -8.92$ & $  0.88$ & - \\
HSCJ023817$-$054555 &  $  0.599$ &  $  1.763$ & $  2.89$ & $  0.85$ & $ 51.19$ & $  0.99$ & $  0.87$ & $ 45.17$ & $  0.92$ & - \\
HSCJ091148$+$041852 & $  0.861^*$ & $  2.627^*$ & $ 10.00$ & $  0.83$ & $-78.45$ & $  2.39$ & $  0.40$ & $ 74.04$ & $  1.24$ & found by \Yattalens \\
HSCJ115252$+$004733 &  $  0.466$ &  $  3.760$ & $  3.63$ & $  1.00$ & $-86.82$ & $  3.28$ & $  0.33$ & $ 19.22$ & $  1.45$ & \cite{MoreEtal17} \\
HSCJ141136$-$010216 & $  0.949^*$ & $  3.021^*$ & $  6.48$ & $  0.56$ & $-52.30$ & $  1.28$ & $  0.74$ & $-47.67$ & $  1.07$ & found by \Yattalens \\
HSCJ144320$-$012538 & $  0.890^*$ & $  1.703^*$ & $  2.76$ & $  0.73$ & $-52.94$ & $  1.68$ & $  0.52$ & $-56.09$ & $  1.09$ & - \\
\end{tabular}
\end{center}
\caption{
Lens candidates with X-shooter spectra. 
We measure the lens and source redshifts as shown in columns 2 and 3.
We highlight the new redshift measurements with $*$ sign.
The best-fit parameter values of lens modeling from \Glee\ is listed in columns 4-9, as illustrated in \fref{fig:glee}.
The corresponding scaled Einstein radius as listed in column 10. 
We allow for an additional external shear component for HSCJ022622$-$042522 and find $\gamma_{\rm ext}=0.15$ and $PA_{\rm ext}=52.27\deg$, due to the presence of a close-by galaxy group.  
}
\label{tab:xshooter}
\end{table*}

\subsection{Lens modeling}

To investigate the lensing nature of the 6 X-shooter lenses, we use the lens modeling software \Glee\  \citep{Suyu&Halkola10,SuyuEtal12}, to fit the lens light and lensed-source components. 
First, we mask out nearby galaxies for each lens.
We model the lens light components using S\'ersic profiles and the lensed-source components using four PSFs, assuming that our candidates have point-like quasar sources.
The best-fit values of S\'ersic index ($n_\text{S\'ersic}$), axis ratio ($q_{\rm light}$) and position angle ($PA_{\rm light}$) are listed in \tref{tab:xshooter}.

\begin{figure*}
\centering
\includegraphics[scale=0.4]{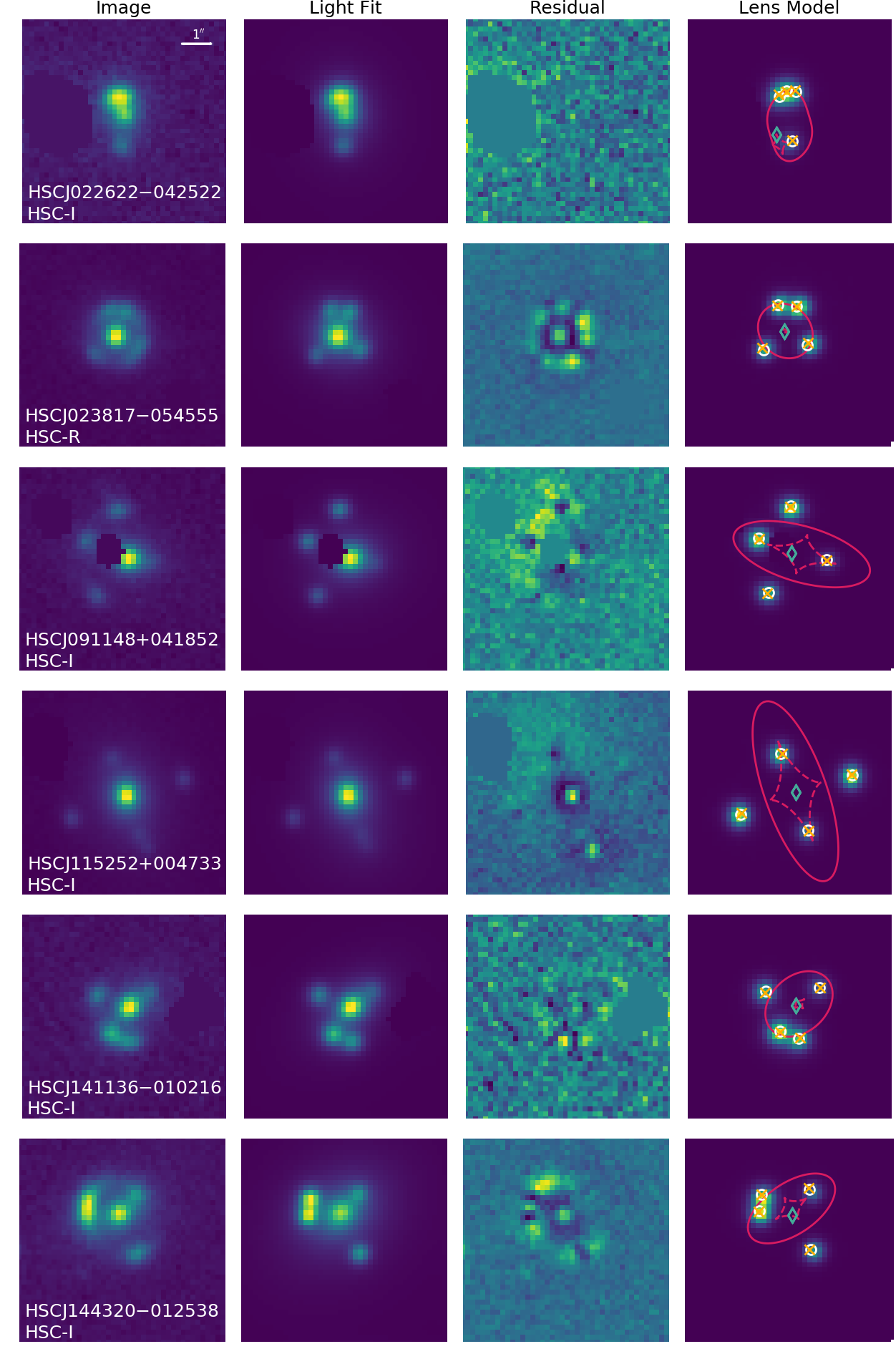}
\caption{
The best fit of lens modeling from \Glee. 
The first column shows the image in the filter with the sharpest PSF. We further mask out nearby objects.
The second column shows the model with a S\'ersic light profile and 4 PSFs, and the residual is shown in the third column.
The fourth column shows the best-fit SIE model. 
The critical and caustics curves are shown in the red solid and dash curves, respectively.
The predicted positions of images and sources are labeled as orange crosses and green diamonds, respectively.
The positions of the 4 PSFs are labeled as white circles.
We impose external shear for HSCJ022622$-$042522 to obtain better modeling, due to the nearby galaxy group.
Each cutout is $7\arcsec\times7\arcsec$.
}
\label{fig:glee}
\end{figure*}

After identifying the positions of four PSFs, we fit the SIE lens model to the four PSF positions.
The reason that we have this examination is to see if the lensed feature can be captured by four simple PSFs. 
The result of \Glee\ is shown in \fref{fig:glee} and the scaled Einstein radii are listed in \tref{tab:xshooter}.
In \fref{fig:glee}, the first column shows the image cutout of the system in the filter with the sharpest PSF. We further mask out the nearby objects.
The second column shows the model with a S\'ersic lens light profile and 4 PSFs, and the residual is shown in the third column.
The fourth column shows the best SIE model. 
The critical and caustics curves are shown in the red solid and dash curves, respectively.
The predicted positions of images and sources are labeled as orange crosses and green diamonds, respectively.
The measured positions of the 4 PSFs are labeled as white circles.
We discuss each object in detail below.
\begin{itemize}
    \item HSCJ022622$-$042522: The lensed images can be well fitted by PSFs. The source is likely to be a quasar. However, we need to impose an additional external shear component to model the image configuration, due to a nearby galaxy group. 
    We also note that the top lensed images are not able to be fitted by single PSF. 
    Therefore this target are more likely to be a quad system.
    The nearby galaxy group results in the substantial difference between $q_{\rm light}$ and $q$. See \fref{fig:gallery}(b). 
    \item HSCJ023817$-$054555: There is evident arc-like residuals, showing that the source is more likely to be a lensed galaxy without AGN. This lens is also found by \Yattalens\ \citep{SonnenfeldEtal18}. Comparing to the lensing parameters from \cite{SonnenfeldEtal19} with $\thetaEin=0.93\arcsec$, $q=0.92$, and $PA=119.8\deg$, the Einstein radius and axis ratio agree well although the $PA$ is offset mostly due to the mass distribution being quite round. 
    The axis ratios and position angles of lens light and lens mass are comparable. 
    \item HSCJ091148$+$041852: The residual may come from the host galaxy of quasar or a galaxy-scale source. 
    The mass profile is more elliptical than the light profile due to some nearby galaxies. 
    \item HSCJ115252$+$004733: The lensed images can be well fitted by PSFs. The mass profile is more elliptical due to the small satellite close to the bottom lensed image. 
    We further compare our lensing parameters to the ones from \cite{MoreEtal17}: $\thetaEin = 4\pi(\frac{\sigma}{c})^2\frac{1}{\sqrt{q}}=3.07\arcsec$ ($\sigma=280$~km/s, $q=0.54$), and $PA=19.1\deg$, which are consistent with the result from \Glee.
    \item HSCJ141136$-$010216: The residual is not prominent. The source could be point-like.  
    The lens mass distribution is rounder, but has the same orientation as the light distribution.
    \item HSCJ144320$-$012538: There is evident arc-like residuals, showing that the source is more likely to be a lensed galaxy without AGN. 
    The orientations of lens light and lens mass distribution are comparable. $q_{\rm light}>q$ is due to imperfect lens light subtraction.
\end{itemize}
We further compare $\thetaEinScaled$ from \Chitah\ and \Glee.
\Chitah, which can rapidly model the lenses, provides good measurements that are within $<8\%$ of the results from the more detailed modeling with \Glee.
We found that three of our X-shooter lenses have well aligned light and mass distribution ($<6\deg$), and two of these three have rounder mass distribution \citep[similar to e.g.,][]{RusuEtal16,ShajibEtal19}. 
Those with $q<q_{\rm light}$ have either near by galaxies or imperfect lens light subtraction.

\subsection{Nature of the sources}

\begin{figure*}
\centering
\includegraphics[scale=0.48]{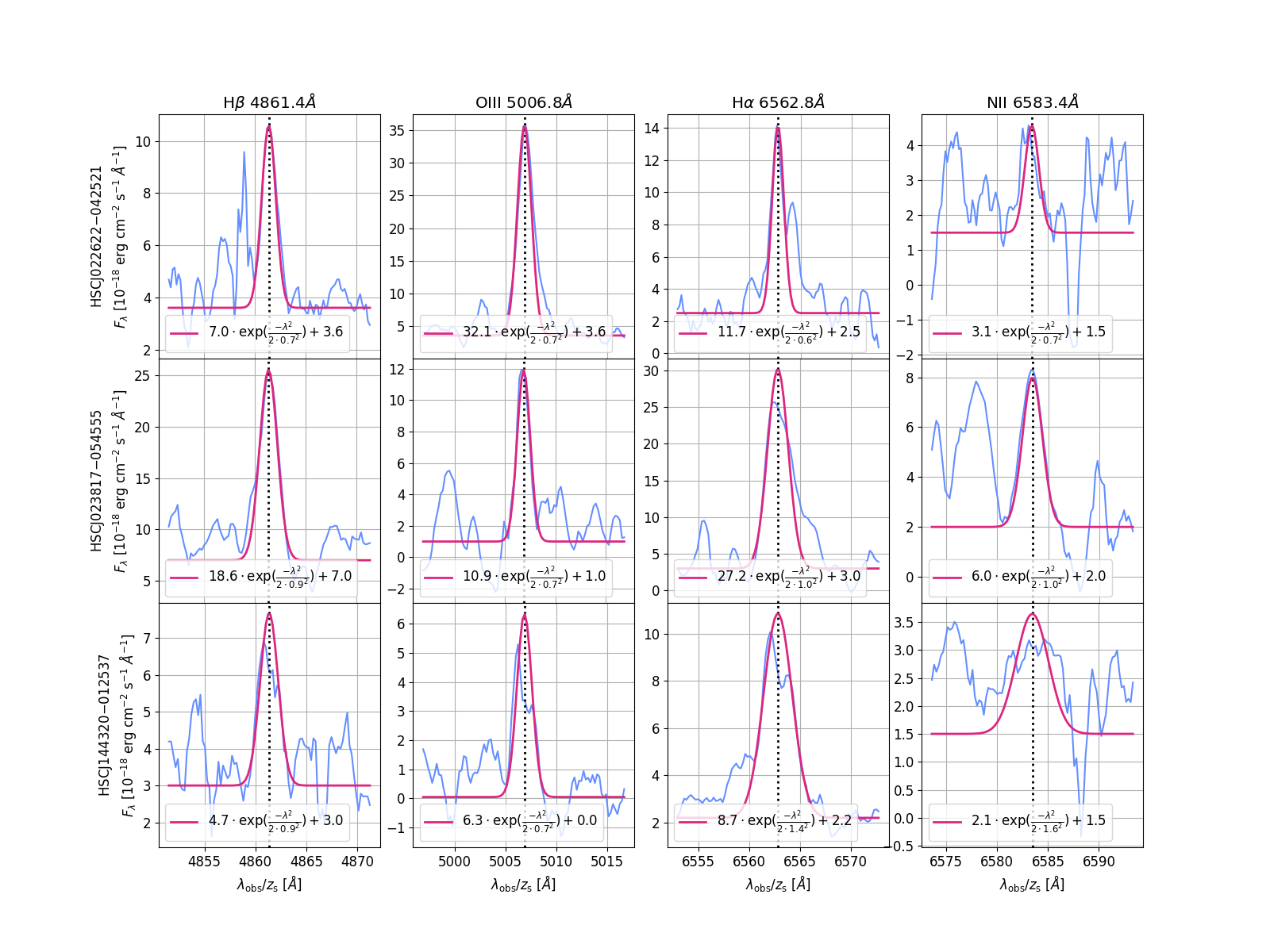}
\caption{The fluxes of emission lines: [{\sc Oiii}], $H_\beta$, [{\sc Nii}], and $H_\alpha$. The lines are detected in the X-shooter NIR arm. We can only obtain upper limits on  [{\sc Nii}]. The red lines show Gaussian fits to the emission lines. 
} 
\label{fig:bpt_sup}
\end{figure*}

\begin{figure}
\centering
\includegraphics[scale=0.5]{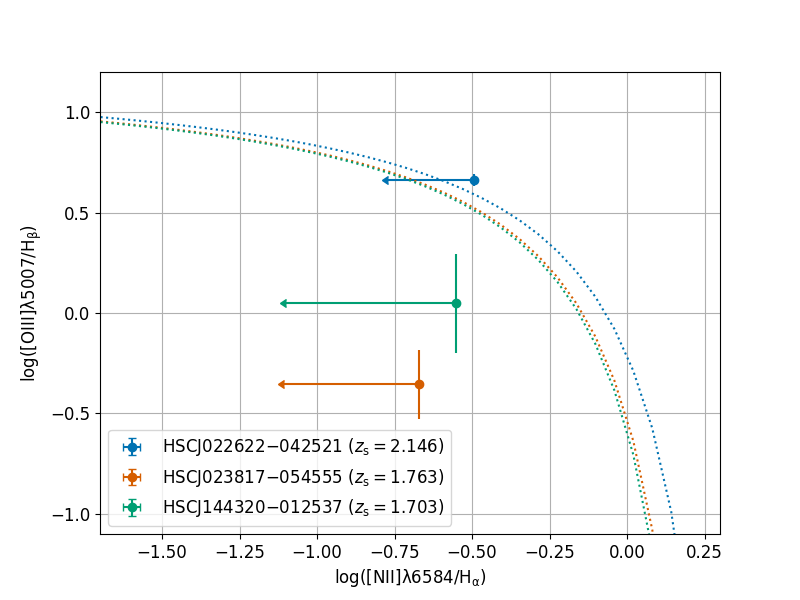}
\caption{
The BPT diagram. The empirical division between star-forming galaxies and AGNs is shown in the dotted curve using \eref{equ:bpt}: below the curve as star-forming galaxies and above the curve as AGNs. 
}
\label{fig:bpt}
\end{figure}

The Baldwin-Phillips-Terlevich (BPT) diagram is commonly used to separate the star-forming galaxy population and AGNs \citep{BaldwinEtal81}.
It allows us to investigate further the nature of the lensed sources. 
We measure the emission-line ratios ([{\sc Oiii}]/H$\beta$ versus [{\sc Nii}]/H$\alpha$) using the spectra as shown in \fref{fig:specz_src}.
The flux of each emission line is fitted by a Gaussian, as shown in \fref{fig:bpt_sup}.
Fortunately, we have three candidates with line detection: HSCJ022622$-$042522, HSCJ023817$-$054555 and HSCJ144320$-$012538, though we can only estimate the upper limit of the [{\sc Nii}]. 
The resulting line ratios on the BPT diagram are shown in \fref{fig:bpt}, and the empirical curve (highlighted in dotted curve) provided from Equation (1) of \cite{KewleyEtal13} has a functional form:
\be
\begin{aligned}
\log([\textsc{Oiii}]/{\rm H}\beta)&= \frac{0.61}{\log([\textsc{Nii}]/{\rm H}\alpha)-0.02-0.1833\zs} \\
&+1.2+0.03\zs.
\end{aligned}
\label{equ:bpt}
\ee
Galaxies below the curve are considered as star-forming galaxies and those above the curve are considered as AGNs.
We notice that only HSCJ022622$-$042522 reaches above the curve, showing that the source is likely to be an AGN. 

\begin{figure}
\centering
\includegraphics[scale=0.6]{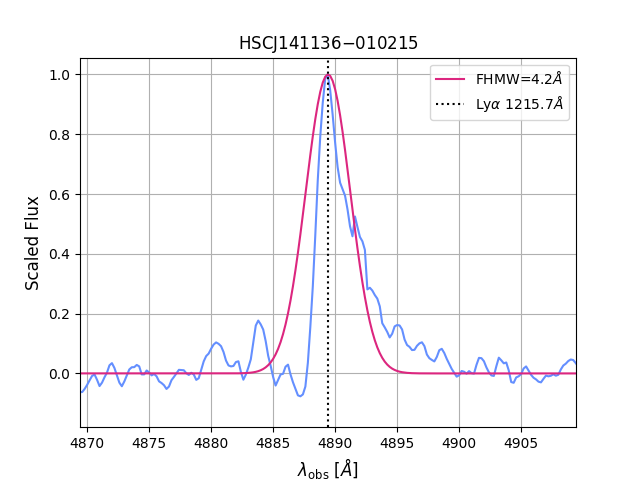}
\caption{
The Ly$\alpha$ emission of HSCJ141136$-$010216. 
The red line show a Gaussian fit.
}
\label{fig:lya}
\end{figure}

For HSCJ141136$-$010216, we detect the Ly$\alpha$ emission only and no other prominent lines, indicative of an AGN, similar to HSCJ115252$+$004733.
Following the method in \cite{MoreEtal17}, we measure the FWHM of Ly$\alpha$ emission to be 4.2\AA\ ($\sim258$\,km\,s$^{-1}$) by fitting a Gaussian (as shown in \fref{fig:lya}). This translates to a velocity width of $252$\,km\,s$^{-1}$ after accounting for the instrumental broadening. 
The Lyman-$\alpha$ emitters (LAEs) have average velocity widths of $264$\,km\,s$^{-1}$ \citep{OuchiEtal10}. Thus, the source of HSCJ141136$-$010216 is most likely to be an LAE, a compact source with finite size rather than an AGN, which is also consistent with the hint of extension seen in the lensed images.

\section{Conclusion and Discussion}
\label{sec:conclusion}
In this work, we present new lens candidates in the HSC survey, selected mainly by \Chitah. We confirm the lens features based on spectroscopic follow-up and lens modeling.
We draw the conclusion as follows
\begin{itemize}
    \item After preselecting objects from either LRG catalogs or QSO catalog, we employ \Chitah\ to classify those within the HSC S16A footprint. 
    We obtain 46 lens candidates with grade larger than 1.5, and 3 of them are previously known lenses which are recovered by \Chitah. 
    \item Including the other two lenses found by \Yattalens\ and one lens in \cite{MoreEtal17}, we obtain X-shooter spectra of 6 objects and confirm them as lenses.
    \item The spectroscopic redshifts of lenses and sources are listed in \tref{tab:xshooter}. 
    We highlight 4 new redshift measurements for both lens and source.
    \item We use \Glee\ to examine the point-like lensed feature of the 6 confirmed lens systems. 
    HSCJ022622$-$042522, HSCJ141135$-$010216 and HSCJ115252$+$004733 are likely to have point-like sources.
    \item We plot the BPT diagram to investigate the nature of the lensed source. 
    HSCJ022622$-$042522 shows that its source is possibly a quasar, though we can only measure the upper limit of the [{\sc Nii}].
    \item We measure the FWHM of Ly$\alpha$ emission of HSCJ141136$-$010216 to be $\sim254$\,km\,s$^{-1}$, showing that it is likely to be a Lyman-$\alpha$ emitter.
    \item As a result of modeling, we found that the lens mass distribution is rounder but well aligned with the lens light distribution, except for those having nearby galaxies or imperfect light subtraction.
\end{itemize}
Though only one possible lensed quasar with spectroscopy available is presented in this work, we note that most of our lenses with X-shooter spectra are high redshift lens systems. These lenses will help us to expand the redshift range for the study of the evolution of lens galaxies.
%


\section*{Acknowledgements}
J.~H.~H.~C.~acknowledges support from the Swiss National Science Foundation (SNSF). S.~H.~S.~thanks the Max Planck Society for support through the Max Planck Research Group.
A.~S. acknowledges funding from the European Union's Horizon 2020 research and innovation programme under grant agreement No 792916, as well as a KAKENHI Grant from the Japan Society for the Promotion of Science (JSPS), MEXT, Number JP17K14250. 
A.~T.~J.~is supported by JSPS KAKENHI Grant Number 17H02868. %
This work was supported by World Premier International Research Center Initiative (WPI Initiative), MEXT, Japan. 
A.~Y.~acknowledges JSPS KAKENHI Grant Number JP25870893.
K.~C.~W. is supported in part by an EACOA Fellowship awarded by the East Asia Core Observatories Association, which consists of the Academia Sinica Institute of Astronomy and Astrophysics, the National Astronomical Observatory of Japan, the National Astronomical Observatories of the Chinese Academy of Sciences, and the Korea Astronomy and Space Science Institute.

The Hyper Suprime-Cam (HSC) collaboration includes the astronomical communities of Japan and Taiwan, and Princeton University.  The HSC instrumentation and software were developed by the National Astronomical Observatory of Japan (NAOJ), the Kavli Institute for the Physics and Mathematics of the Universe (Kavli IPMU), the University of Tokyo, the High Energy Accelerator Research Organization (KEK), the Academia Sinica Institute for Astronomy and Astrophysics in Taiwan (ASIAA), and Princeton University.  Funding was contributed by the FIRST program from Japanese Cabinet Office, the Ministry of Education, Culture, Sports, Science and Technology (MEXT), the Japan Society for the Promotion of Science (JSPS),  Japan Science and Technology Agency  (JST),  the Toray Science  Foundation, NAOJ, Kavli IPMU, KEK, ASIAA,  and Princeton University.

The Pan-STARRS1 Surveys (PS1) have been made possible through contributions of the Institute for Astronomy, the University of Hawaii, the Pan-STARRS Project Office, the Max-Planck Society and its participating institutes, the Max Planck Institute for Astronomy, Heidelberg and the Max Planck Institute for Extraterrestrial Physics, Garching, The Johns Hopkins University, Durham University, the University of Edinburgh, Queen's University Belfast, the Harvard-Smithsonian Center for Astrophysics, the Las Cumbres Observatory Global Telescope Network Incorporated, the National Central University of Taiwan, the Space Telescope Science Institute, the National Aeronautics and Space Administration under Grant No. NNX08AR22G issued through the Planetary Science Division of the NASA Science Mission Directorate, the National Science Foundation under Grant No. AST-1238877, the University of Maryland, and Eotvos Lorand University (ELTE).
This paper makes use of software developed for the Large Synoptic Survey Telescope. We thank the LSST Project for making their code available as free software at http://dm.lsst.org.
Based in part on data collected at the Subaru Telescope and retrieved from the HSC data archive system, which is operated by the Subaru Telescope and Astronomy Data Center at National Astronomical Observatory of Japan.
This work is based in part on observations collected at the European Southern Observatory under ESO programme 099.A-0220.


\bibliographystyle{aa}
\bibliography{SuGOHI-q}


\end{document}